\documentclass{sig-alternate-05-2015}
\usepackage{subfigure}
\usepackage{algorithm}
\usepackage{algorithmic}
\usepackage{multirow}
\usepackage{graphicx}
\usepackage{epstopdf}

\begin{document}

\setcopyright{acmcopyright}

\doi{10.475/123_4}

\isbn{123-4567-24-567/08/06}

\conferenceinfo{PLDI '13}{June 16--19, 2013, Seattle, WA, USA}

\acmPrice{\$15.00}

%
\conferenceinfo{WOODSTOCK}{'97 El Paso, Texas USA}

\title{Large Scale Product Graph Construction for Recommendation in E-commerce}

%
%
%
%
\numberofauthors{1} 
%
\author{
\alignauthor {Xiaoyong Yang${}^\dag$ \hspace*{0.3cm} Yadong Zhu${}^\dag$\titlenote{Corresponding Author} \hspace*{0.3cm}Yi Zhang${}^\ddag$ }\hspace*{0.3cm}Xiaobo Wang${}^\dag$ \hspace*{0.3cm}Quan Yuan${}^\dag$\\
\affaddr{\dag~Alibaba Inc, Beijing, China} \\
\affaddr{\ddag~University of California, Santa Cruz, USA}\\
\email{\{xiaoyong.yxy, edgewind.zyd, yongshu.wxb, yuanquan.yq\}@alibaba-inc.com} \\
\email{\{yiz\}@soe.ucsc.edu }
}



\maketitle
\begin{abstract}

Building a recommendation system that serves billions of users on daily basis is a challenging problem, as the system needs to make astronomical number of predictions per second based on real-time user behaviors with O(1) time complexity. Such kind of large scale recommendation systems usually rely heavily on pre-built index of products to speedup the recommendation service so that the waiting time of online user is un-noticeable. One important indexing structure is the product-product index, where one can retrieval a list of ranked products given a seed product. The index can be viewed as a weighted product-product graph.

In this paper, we present our novel technologies to efficiently build such kind of indexed product graphs.
In particular, we propose the Swing algorithm to capture the substitute relationships between products, which can utilize the substructures of user-item click bi-partitive graph. Then we propose the Surprise algorithm for the modeling of complementary product relationships, which utilizes product category information and solves the sparsity problem of user co-purchasing graph via clustering technique.
Base on these two approaches, we can build the basis product graph for recommendation in Taobao.
The approaches are evaluated comprehensively with both offline and online experiments, and the results demonstrate the effectiveness and efficiency of the work.

\end{abstract}

\category{H.2.8}{Database Management}{Database applications}--{Data mining}

\terms{Algorithms, Design, Experimentation}

\keywords{Product Graph, E-commerce, Recommender System}

\section{Introduction}

\begin{figure*}
\centering
\epsfig{file=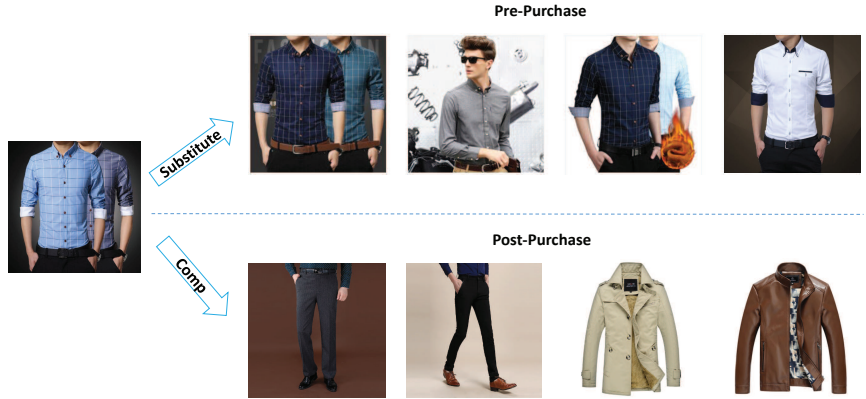, width=4.5in}
\caption{Application example of different relationships between products.}\label{demo}
\end{figure*}

As billions of users leverage e-commerce sites to fulfill their purchasing needs while saving time and voiding crowds, and how to better serve online customers is becoming an increasingly important problem. Meanwhile large e-commerce sites such as Taobao and Amazon often provide billions of products for consumers to choose. Recommender systems can help consumers to select the right products to meet their tastes and special needs, and play key role in modern e-commerce \cite{aicf}.
Existing recommendation approaches can be mainly divided into two categories: Collaborative Filtering approaches (CF) \cite{Linden:2003:ARI, Wang:2011:URP} and Content-Based approaches (CB) \cite{ Lops:2011, Mooney:2000:CBR, Pazzani:1997}. There also exist other approaches such as hybrid methods that combines both, or context-aware methods \cite{Adomavicius:2008:CRS, DBLP:journals/aim/AdomaviciusMRT11}, which considers user's current context information to make recommendation more reasonable.

Besides the detailed recommendation approaches, another important problem is understanding and capturing relationships between products, which is the basis of modern e-commerce recommender systems \cite{McAuley:2015:INS}. We can view product relationship graph as a re-built index of products, which can return back a list of ranked products given a seed product. This index can dramatically speedup the recommendation service so that online user waiting time is un-noticeable.

There are two very important types of relationships between products: \textit{substitute} and \textit{complementary}.  \textit{Substitute} products are those that are interchangeable such as these shirts shown in Figure~\ref{demo}, while \textit{complementary} products are those that might be purchased in addition such as shirts and trousers \cite{McAuley:2015:INS}.
Different contextual environments may have different requirements or implications about recommendation relevance, thus different relationship graphs are needed to speed up the recommendation. For example, at different stages of a user session, the user's preferences can be very different. As illustrated in figure~\ref{demo}, when a user focuses on shirts and has not purchased any shirt in the session, the user may prefer to get \textit{substitute} products recommended for him/her to compare with. Once a purchase is made, substitute products becomes less relevant, and some \textit{complementary} products such as trousers or jackets would be more attractive and relevant.

Recently some researchers have recognized the importance of substitutes and complements, \cite{McAuley:2015:INS} utilizes the text of product reviews and descriptions to infer the relationships between products via a supervised approach. However, the existing approach doesn't apply to a huge recommendation system like Taobao's, because the text of product reviews and descriptions are huge and very noisy for billions of products from millions of sellers, thus the supervised approach based on text is not cost effective.

On the other hand, Taobao has an enormous amount of real user behavior data from billions of customers, which are much stronger and reliable signals for capturing product relationships than text data. Thus we focus on utilizing the user behavior data directly to construct the product graph via an unsupervised approach similar to similarity based collaborative filtering.

When we try to adapt traditional approaches such as item-item CF with local similarities \cite{Sarwar:2001:ICF, Linden:2003:ARI}, we face the following challenges:
\begin{itemize}
\item \textit{Accuracy}: The traditional local similarity calculations do not consider any inner structure of user behavior graph, which has been shown to be useful for other link prediction problems. Thus the prediction accuracy is limited.
\item \textit{Robustness}: User behavior data (i.e.~user-item click graph) contain many noisy, casual or accidental clicks, which could affect the reliability of predictions.
\item \textit{Sparsity}: Though the user behavior data in Taobao is huge, the ratio of user purchase is relatively small. User co-purchasing data is very sparse, thus capturing complementary relationships is very difficult.
\item \textit{Direction}: The complementary relationship is asymmetry. We also need to consider the direction of relationship on the basis of co-purchasing graph.
\item \textit{Scalability}: The computation complexity of traditional approaches grows with both the number of customers and products. Given billions of users and products in Taobao, scalability is a major challenge.
\end{itemize}

In this paper, we propose a new algorithm called Swing, which can utilize the inner structure--\textit{Swing} of user-item behavior graph, to construct substitute product graph. Swing is a quasi-local structure and is designed to be much more stable than a single edge used in traditional CF approaches. It provides much more reliable calculation propagation over a user-item bi-partite graph, and helps to reduce the influence of noisy user behavior data.
Then we propose a new algorithm called Surprise to construct complementary product graph. Surprise algorithm solves the sparsity problem on the user co-purchasing graph by utilizing products' category information and product clusters constructed based on Swing algorithm. Furthermore, it considers the time sensitivity and temporal order of co-purchased products.
Both methods are implemented for parallel running with commonly used large scale distributed computing framework such as Map-Reduce or Spark, thus scalability is not a problem.
Based on that, we can build the product substitute graph and product complementary graph in Taobao, which provide basic indexing service to generate candidate products for further recommendation ranking modules.

To evaluate the effectiveness of our approaches, we conduct extensive experiments with both offline data and online user studies. The offline experimental results on a large dataset demonstrate the proposed approaches can significantly outperform an existing well tuned baseline CF method in Precision, Recall and MAP metrics. Online experimental results with real world e-commerce users on Taobao also demonstrate the proposed approaches lead to significantly higher CTR (click through rate), CVR (conversion rate) and PPM (Payment Per Thousand Impressions). The efficiency of the new approaches on offline running time is also analyzed and demonstrated.

The main contribution of this papers are:
\begin{enumerate}
\item A new efficient and effective algorithm (i.e.~Swing) that utilizes the quasi-local structure information of user behavior graph. It provides a more reliable calculation propagation and eliminates the effect of noisy data. Based on Swing, we can build more reliable similar relationships between products.
\item A new efficient and effective algorithm (i.e.~Surprise) that utilizes product category information and clustering technique. It solves the sparsity problem, thus the complementary relationships between products are more reliable and reasonable when using the Surprise algorithm.
\item An efficient large scale industrial implementation for the proposed approaches via parallelization.
\item An empirical verification of the effectiveness and efficiency of the new approaches via comprehensive offline and online experiments and detailed analysis.
\end{enumerate}

The rest of the paper is organized as follows. We introduce the Swing algorithm for substitute relationships in Section 2. We then describe the Surprise algorithm for complementary relationships in Section 3. Section 4 presents the experimental results, Section 5 describes the related work, and Section 6 concludes the paper.

\section{Swing Algorithm for Substitute Relationships}

\begin{figure*}[!htb]
 \centering
  \subfigure[]{
    \label{fig:ts:a} 
    \includegraphics[width=0.4 \textwidth]{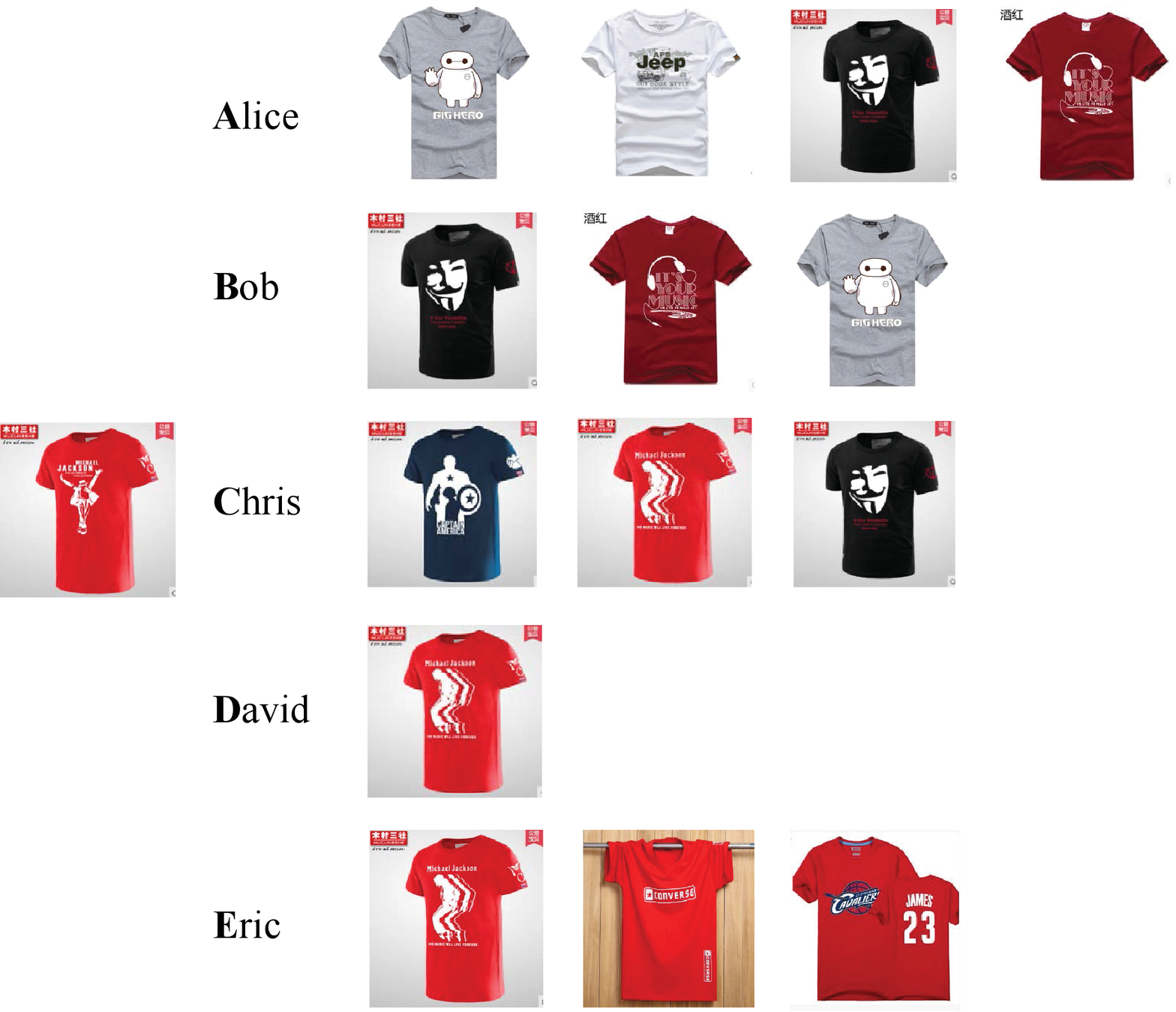}}
  \subfigure[]{
    \label{fig:ts:b} 
    \includegraphics[width=0.26 \textwidth]{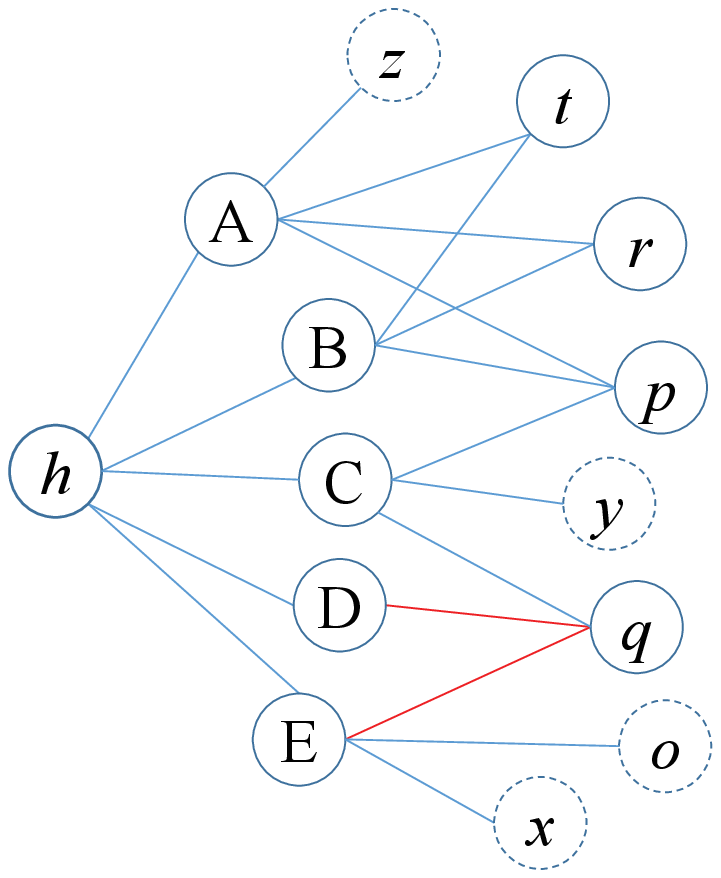}}
  \caption{An Example of Swing score computing.}
\end{figure*}

Computing similarity between two items is central to the task of building substitute graph. There are many similarity measures that only require local information and those methods are usually computationally efficient and can be applied to extremely large graphs. Frequently used methods in recommendation are shown as follows \cite{Sarwar:2001:ICF, zhou2009predicting}.

\emph{Vector Cosine Similarity}
$$w_{i,j}=\frac{U_{i}\bigcap U_{j}}{\|U_{i}\|\times\|U_{j}\|}$$
where $U_{i}$ is the set of users that have clicked item $i$.

\emph{Jaccard Coefficients}
$$w_{i,j}=\frac{U_{i}\bigcap U_{j}}{U_{i}\bigcup U_{j}}$$

\emph{Correlation-Based Similarity}

$$w_{i,j}=\frac{\sum_{u \in U_{i}\bigcap U_{j}}{ (r_{u,i}-\overline{r_{i}})( r_{u,j}-\overline{r_{j}}) } }
{\sqrt{\sum_{u \in U_{i}\bigcap U_{j}}{{(r_{u,i}-\overline{r_{i}})^2} }}\sqrt{\sum_{u \in U_{i}\bigcap U_{j}}{{(r_{u,j}-\overline{r_{j}})^2} }}}$$
where $r_{u,i}$ is the rating of user $u$ on item $i$, and $\overline{r_{i}}$ is item $i$'s average rating. The similarity between item $i$ and item $j$ is measured by computing \emph{Pearson Coefficients}.

Most traditional methods focus on item-user-item path when calculating neighborhood strength, with item based normalization to penalize popular items. These methods do not utilize the other inner structures such as user-item-user on the user behavior graph, thus the prediction accuracy is limited.

For example, suppose there are five users (Alice, Bob, Chris, David and Eric) and all of them are looking for T-shirts on Taobao. As shown in figure~\ref{fig:ts:a}, each row represent a user and items the user clicked. Alice is choosing T-shirt for her boyfriend. Bob wants to buy a popular T-shirt for himself. Chris loves t-shirts of a special brand named \emph{Mucunsanshe}. David is a big fan of Michael Jackson, and only clicks one Michael Jackson T-shirt. While Eric is looking for T-shirts in red color. All of them have clicked item $h$. The click information can be summarized as figure~\ref{fig:ts:b}, where each capitalized letter represents a person(corresponding to five users above, we use A,B,C,D,E for short) and each lower case letter represents a clicked product.
For simplicity and without loss of generality, we only show items in the same category: \emph{T-shirts}. Users may click any objects. For example, after buying T-shirts, Alice may view and clicks dresses for herself, and David may click headphones, since he is a music enthusiast.

In this case, we have $|U_h|=5$, $|U_t|=15$, $|U_p|=40$, $|U_q|=60$ and $|U_z|$=4, which means there are totally 5 users that have clicked item $h$, and 15 users have clicked item $t$, etc. If we rank neighboring items for $h$ using cosine similarity, the result would be t > z > p > q. However, we can see that $z$ is not very similar to $h$. Besides the number of common-neighbors, it only depends on the degree of each item node on the denominator(namely, how many users have clicked each item), which could be affected by many factors and has big uncertainty. Noting that the $CF$ score of $z$ could be greater than $p$ and $q$, as it is less popular.

\subsection{Swing Algorithm}\label{ss:sw}

Since our user click data is noisy, with many accidental or random clicks, we need a node-proximity measure which will consider more robust inner-network structural information. Stable network structure has been studied in link prediction field, e.g. users tend to form triangles in social networks \cite{Kossinets:2006p, LichtenwalterC12}.
Considering bipartite-networks in E-commerce, when there is only one user who clicks $h$ and $i$ together, it is more likely to be a coincidence. However, if two users both clicked $h$ and $i$, the relation is much stronger. Thus for each seed product, considering the local graph that includes all users who clicked the seed product and all products clicked by those users, we call each $user\mbox{-}item\mbox{-}user$ network structure on the local graph as a $swing$. For example, besides the seed item $h$, both $D$ and $E$ also clicked item $q$, so $[D, q, E]$ is a $swing$. while $[A, z], [C, y], [E, o], [E, x]$ are just single edges that are not part of any swing.

If there are many $swing$s between a user pair $<u,v>$, it usually indicates a wide range of products could meet their requirements. And the relation among these $swing$s is less intensive. Therefore, we weigh each $swing$ through the total number of $swing$s formed between each user pair. The definition of $swing$ score is given below.

\begin{equation}\label{eq:sw1}
s(i,j)=\sum_{u \in U_{i}\bigcap U_{j}}\sum_{v \in U_{i}\bigcap U_{j}}\frac{1}{\alpha + |I_{u}\bigcap I_{v}|}
\end{equation}

where $U_{i}$ is the set of users who have clicked item $i$, and $I_{u}$ is the set of items that clicked by user $u$. $\alpha$ is a smoothing coefficient.

Consider the example in figure 2 and how to compute the similarity between item $h$ and other items. Without loss of generality, let $\alpha=1$. $[A, p, B]$ is a swing, while there are 3 swings on $[A, B]$ with $[A, t, B]$, $[A, r, B]$ and $[A, p, B]$. Therefore, the swing score $p$ contributed by $[A, B]$ is $\frac{1}{(1 + 3)}$. $[B, p, C]$ is a swing, and there is only one swing on $[B, C]$, so the swing score of $p$ contributed by [B, C] is $\frac{1}{(1 + 1)}$. Similarly, $p$ also gets $\frac{1}{(1 + 1)}$ from $[A, C]$, as $[A, p, C]$ is a swing. The total swing score for $p$ is:

$$swing(h,p) = \frac{1}{4} + \frac{1}{2} + \frac{1}{2} = 1.25$$
Similarly, we have:
$$swing(h,q) = \frac{1}{2} + \frac{1}{2} + \frac{1}{2} = 1.5$$
$$swing(h,t) = swing(h,r) = \frac{1}{2} = 0.25$$

Therefore, we have $ q > p > r = t > others$. Item $q$ is a Michael Jackson T-shirt with brand \emph{Mucunsanshe} and red color, which is very similar to item $h$. With $swing$, we are able to rank item $q$ up to the first place. At the same, all items connected with single edge are ranked to the bottom in a strict way.

For each $swing$ structure, we further add user weighting factor similar to the well-known Adamic/Adar algorithm \cite{Liben-Nowell:2003:LPP} to penalize active users. The more items a user clicks, the smaller weight it gets. The final $swing$ score is:

\begin{equation}\label{eq:sw2}
s(i,j)=\sum_{u \in U_{i}\bigcap U_{j}}\sum_{v \in U_{i}\bigcap U_{j}}w_{u}\cdot w_{v}\cdot\frac{1}{\alpha + |I_{u}\bigcap I_{v}|}
\end{equation}

where
$$w_{u} = \frac{1}{\sqrt{|I_{u}|}}, w_{v} = \frac{1}{\sqrt{|I_{v}|}}$$
The detailed procedure to calculate swing score is described in Algorithm~1.

Noting that in this work, we mainly focus on the similarity computing based on user's clicking behavior only, other factors such as time decaying could also be incorporated into each method. Furthermore, extensive closely related clicks often happen in the same user session in Taobao, where time interval shows small impact.

Assuming the total number of items is $T$, the average node degree of an item is $N$, and the average node degree of a user is $M$, the traditional item-item similarity based CF approaches have a time complexity of $O(T\cdot N \cdot M)$. The time complexity of \textit{Swing} is $O(T\cdot N^2 \cdot M)$, which is higher due to the consideration of inner network structures on user behavior data.

\begin{algorithm}[t]\label{alg:swing}
\renewcommand{\algorithmicrequire}{	\textbf{Input:}}
\renewcommand{\algorithmicensure}{	\textbf{Output:}}
\caption {\textbf{Swing Algorithm for Substitute Relationships}}
\begin{algorithmic}[1]
\REQUIRE~~\\
User and Item index table: $\textbf{U}$, $\textbf{I}$, \\
Smoothing coefficient: $\alpha$
\ENSURE Substitute item list for $item_i$
\FOR{$\text{each}~u \in U_i$}
\STATE $w_u=\frac{1}{\sqrt{|I_{u}|}}$
\FOR{$\text{each}~v \in {U_i \setminus u}$}
\STATE $w_v=\frac{1}{\sqrt{|I_{v}|}}$
\STATE $k=|I_u \bigcap I_v|$
\FOR{$\text{each}~j \in {I_u \bigcap I_v}$}
\STATE $Swing[j]=w_u*w_v*\frac{1}{\alpha+k}$
\ENDFOR
\ENDFOR
\ENDFOR
\STATE \textbf{return} $\textbf{Swing}_{i}=({Swing}_{i}[1],\cdots, {Swing}_{i}[n])$.
\end{algorithmic}
\end{algorithm}

\subsection{Parallelization Implementation Framework}

\begin{figure*}
\centering
\epsfig{file=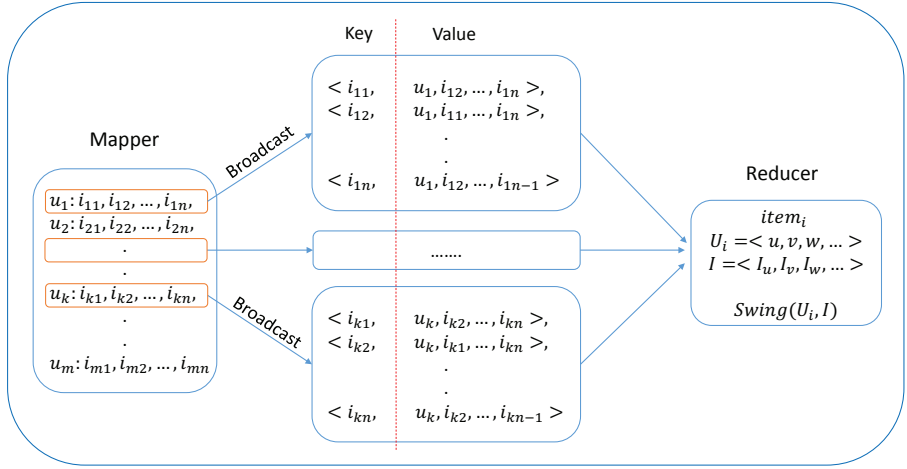, width=4.8in}
\caption{Parallelization implementation of Swing via MR framework}\label{fig:mrp}
\end{figure*}

In the Taobao recommendation system, we developed parallel implementation of \textit{Swing} on a Map-Reduce programming framework in our own distributed platform of Open Data Processing Service\footnote{https://www.aliyun.com/product/odps/}.
The detailed procedure is presented in fig~\ref{fig:mrp}.

Specifically, for the original input data of user click matrix, each row is a clicked item list of a specific user. At the mapper stage, for each item clicked by user $i$ in one row, we make a neighborhood broadcasting through $u_i$, and output the key-value pairs as $<i_{i1}, u1~i_{12}~\cdots~i_{1n}>$. Then at the Reducer stage, we collect the user set $U_i$ who clicked $item_i$, and the item set clicked by each user in $U_i$. Finally \textit{Swing} is calculated to compute the most similar items for $item_i$ as described in Algorithm~1.

\section{Surprise Algorithm for Complementary Relationships}

In this section, we propose a comprehensive framework to find complementary products by mining user purchasing data.
When a user has just bought a T-shirt, it would be unappropriate to recommend T-shirts anymore. Instead, shorts and shoes may be better candidates. If he has just bought a mobile phone, showing accessories such as phone shells and portable power devices are more reasonable, since the user is no longer interested in mobile phones. This important product-product relationship is called \emph{complementary} relationship, which links a seed product with a related product that the user is likely to buy additionally or together.

To find complementary relationships, we borrow the idea proposed in \cite{Wang:2011:URP} and use purchasing data to find candidate products. Furthermore, we need to consider time sensitivity of products \cite{Koren:2010:CFT, Xiang:2010:TRG, Rendle:2010:FPM, Wang:2013:OME} and data sparsity problem, especially in Taobao scenario. First, we introduce a continuous time decay factor to the behavioral relevance, which has been shown to improve accuracy. Then, instead of using content information, which exists in very high dimensional space and differs greatly across categories, we introduce a clustering method based on user click data to solve the problem of data sparsity. Note that implicit feedback data such as user clicks is actually not so sparse, although user purchasing data is sparse, because users normally browse and click lots of items and only buy much less.
We call this framework with the name of \textit{Surprise} to aim at helping users to find closed related items of different categories that users would be unaware.

\subsection{Relevance of Category-level}

To find the most related products to the seed item user purchased, firstly we try to find the most related categories on products' taxonomy, which is usually used in e-commerce systems. This removes a vast majority of cases derived from user purchase behaviors.

By mapping each item to its category, we get a user-category matrix. Then standard collaborative filtering techniques could be used to compute the relevance between categories.
The probability of $c_{j}$ is a related category for $c_{i}$ is defined as follows:

\begin{equation}\label{eq:spc}
\theta_{i,j} = p({c}_{i,j}|c_{j}) = \frac{N(c_{i,j})}{N(c_j)}
\end{equation}
where $N(c_j)$ is the total purchases in category $c_{j}$, and $N(c_{i,j})$ is the number of times $c_{j}$ is purchased after $c_{i}$.

Supposing that the category lists obtained from equation~\ref{eq:spc} for $c_i$ is $[c_{j_1}, c_{j_2},... ,c_{j_m}]$, with a descending posterior probability list
of $[\theta_{i,j_1}, \theta_{i, j_2}, ..., \theta{i, j_m}]$. Roughly take top percentage or a fixed number of candidates as
relevant categories for $c_i$ is not acurrate, since category varies greatly from each other. Instead, we compute a \textit{relative drop} score for each category $c_{j,k}$.
\begin{equation}\label{eq:drop}
\eta_{k} = \frac{(\theta_{i,j_{k+1}} - \theta_{i, j_k})}{\theta_{i, j_k}}
\end{equation}
Categories ranked before the \textit{maximum relative drop point} are selected as top-related categories for $c_i$, and we denote these categories as $\Gamma(c_i)$.

\begin{figure*}[!htb]
 \centering
  \subfigure[]{
    \label{fig:top:a} 
    \includegraphics[width=0.42 \textwidth]{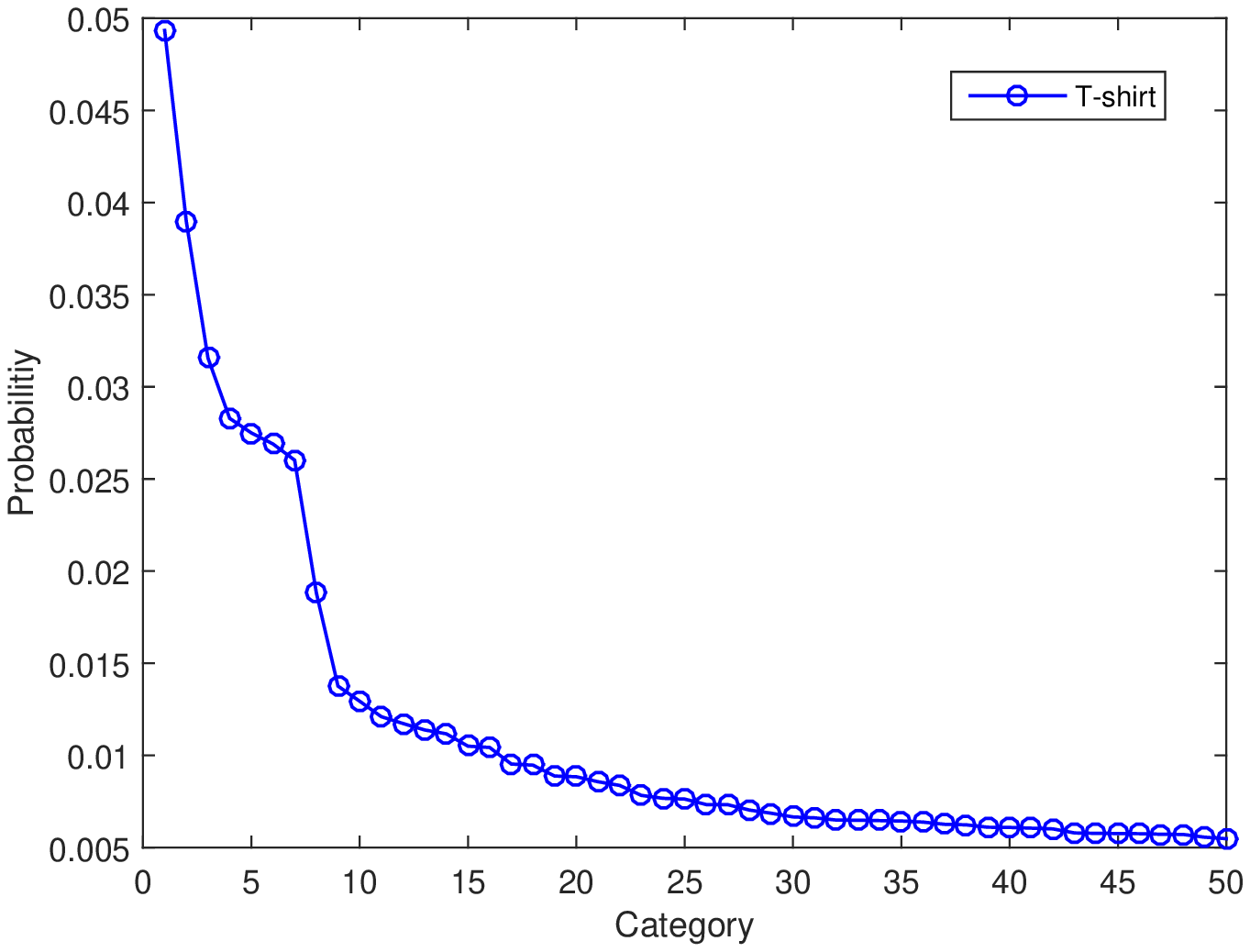}}
  \subfigure[]{
    \label{fig:top:b} 
    \includegraphics[width=0.42 \textwidth]{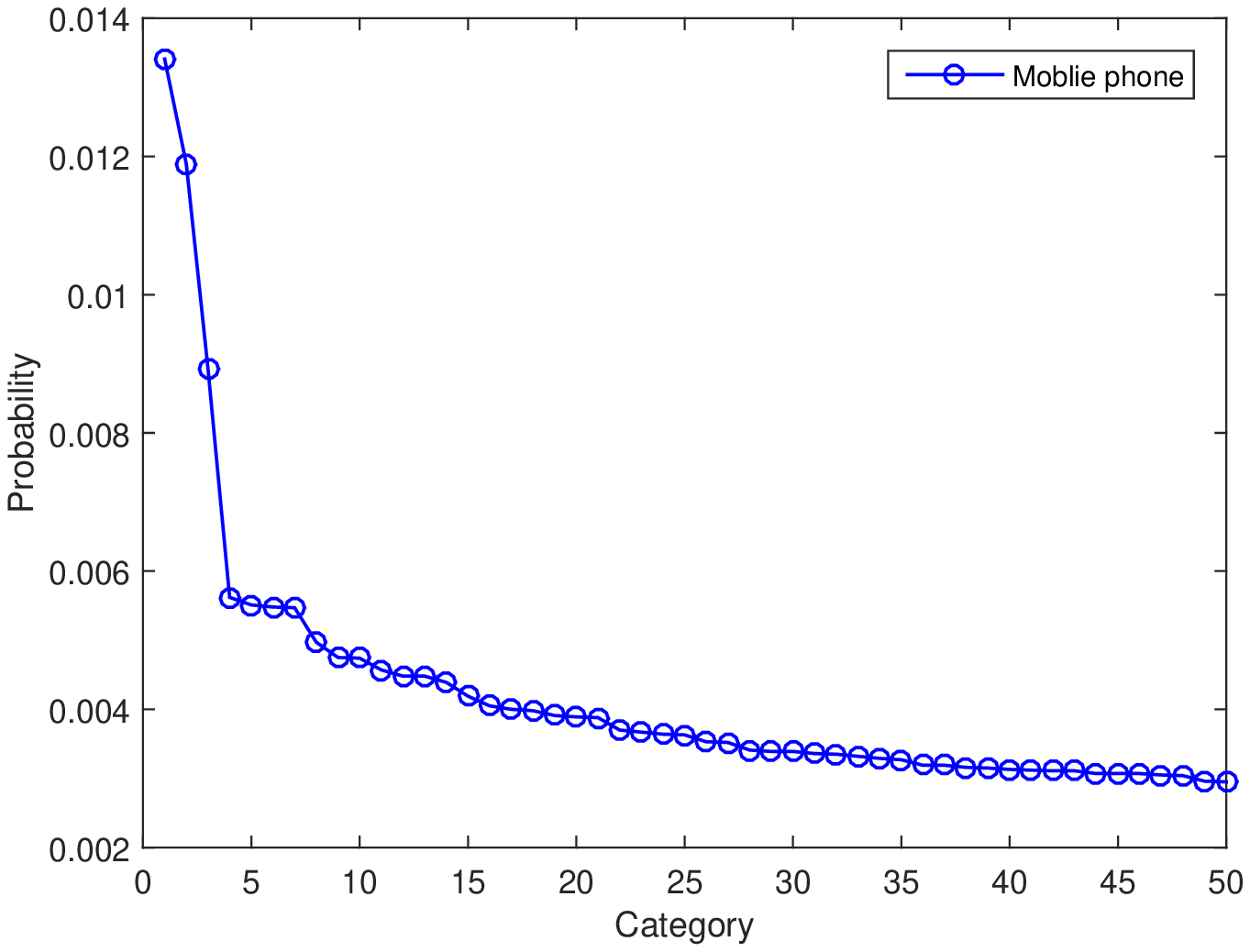}}
   \caption{The examples of Top related categories selection via Max\_drop. }\label{fig:top}
\end{figure*}

The probabilities of related categories for \textit{T-shirt~(male)} and \textit{Mobile Phone} are shown in figure~\ref{fig:top:a} and figure~\ref{fig:top:b} respectively.
From the two sub-figures, we can see there are change points in relevance distribution. By taking items before the maximum relative drop point, we get 8 top-related categories for \textit{T-shirts~(male)}: \textit{Casual Pants}, \textit{Jeans},
\textit{Jacket}, \textit{Shirt}, \textit{Low Shoes}, \textit{Sweater}, \textit{Wei clothing}, \textit{Cotton-padded Coat}, and 3 tightly
related categories for \textit{Mobile phone}: \textit{Phone shell}, \textit{Phone membrane} and \textit{Portable power source}.

\subsection{Relevance of Product-level}

For each related category, we compute the relevance score for items in the category, where standard item-item based collaborative filtering techniques could be applied. In our system, we add the constraint that a candidate related item $j$ should be purchased after item $i$. The \textit{order} is very important in modeling \emph{complementary}. For example, it is reasonable to recommend portable power devices when a user has just bought a cell phone, while it would be unappropriate if we recommend cell phones for him after a portable power device is purchased. Then the related score is defined as follows.

\begin{equation}\label{eq:spi}
s_1(i,j)=\frac{\sum_{u \in {U_i \bigcap U_j}}{1/(1+|t_{ui}-t_{uj}|)}}{{\parallel U_i \parallel} \times {\parallel U_j \parallel}}
\end{equation}
where $c_j \in \Gamma(c_{i})$ and $t_{uj} \ge t_{ui}$. In Equation~\ref{eq:spi}, we add a time decay factor in the numerator. If the time interval between purchase of item $i$ and item $j$ is long, it will be less likely to exist a strong relationship between them.

\subsection{Relevance of Cluster-level}

Supposing that Bob buys jeans $j$ after a T-shirt $i$, we are not sure that $j$ is a good match to $i$. While if there are several users who have bought $j$ after $i$, we'll be more confident that $j$ is very likely to be related to $i$.
We also introduce a threshold on the co-occurrence number of $(i,j)$ when selecting candidate items. For example, when we make recommendation after the purchase of item $i$, we only take the candidate items with $Co(i,j) > \gamma$, which can be viewed as a kind of behavior confidence. However, this constraint will further bring the problem of data sparsity, since the user co-purchasing data is very sparse.

Considering a different situation, where Bob bought $j$ after $i$ and David bought $k$ after $i$, meanwhile item $j$ and item $k$ are very similar, for instance they are both blue levis jeans. This kind of co-purchasing is informative about product relationships. This motivates us to also calculate the relevance score for items at the cluster level, which help to alleviate the data sparsity problem of co-purchasing at item level.

we attempt to project similar items into the same cluster by utilizing the similar items graph constructed by the \textit{Swing} algorithm in the previous section. Then we compute the co-purchasing relevance at cluster-level.

\subsubsection{Clustering using label propagation}

Traditional clustering methods such as k-means and density based methods are not feasible in Taobao scenario, as scalability is an issue with billions of products. Inspired by the community detection method proposed in \cite{PhysRevE.76.036106}, we perform a similar label propagation process in the substitute graph created before.

For each item, top-similar items computed by $Swing$ is added as its neighbors, with a directed link pointing from a similar item to the seed item. The weight of the edge $e_{j,i}$ is set to the similarity score $swing_{ij}$. Note that this weight is not necessarily symmetric. Initially, the \textit{label} for each item node is setting to its node-id.
The $p(\cdot)$ denotes the current probability of the corresponding \textit{label} that a neighbor node belongs to. For each item node, we consider all its neighbors when updating the label probability based on the edge weight, then we choose the label with maximal probability value for the item node based on a certain probability (i.e. $random()> \beta$).

We implement this algorithm using a large scale graph computing framework in the Aliyun platform\footnote{www.aliyun.com}, and it converges after about ten iterations. Finally items with the same label are grouped into the same cluster. The algorithm is proved to be a very fast yet effective clustering method. It takes only \textit{15 minutes} to cluster billions of items, and the detailed algorithm is presented in Algorithm~2.

\begin{algorithm}[t]\label{alg:cluster}
\renewcommand{\algorithmicrequire}{	\textbf{Input:}}
\renewcommand{\algorithmicensure}{	\textbf{Output:}}
\caption {\textbf{Clustering using label propagation}}
\begin{algorithmic}[1]
\REQUIRE~~\\
Similarity graph $G(V, E)$, \\
item node $x \in V$ and its neighbors $\Gamma(x)$,\\
damping factor $\beta$=0.25
\ENSURE Unique label $L(x)$
\STATE \text{init}~$L(x)=x$
\FOR{$t=1,...,n$}
\FOR{$\text{each}~x \in V$}
\STATE \text{init}~$p[L[y]]=0,~y \in \Gamma(x)$
\FOR{$\text{each}~y \in \Gamma(x)$}
\STATE $p[L[y]]+=e_{y,x}$
\ENDFOR
\IF{$random()>\beta$}
\STATE \text{Set} $L(x)=k, \text{where}~p[k]=max(p[1:m])$
\ENDIF
\ENDFOR
\ENDFOR
\STATE \textbf{return} $\textbf{L}$.
\end{algorithmic}
\end{algorithm}

\subsubsection{Cluster-level relevance}
After clustering items into different groups, we compute the relevance score at the cluster level. Let $L(i)$ be the cluster to which item $i$ belongs. Then

\begin{equation}\label{eq:s2}
s_2(i,j)=s_1(L(i),L(j))
\end{equation}
where $s_1$ is computed as equation~\ref{eq:spi} described before.
That is, we compute a relevance score that purchases of item cluster $L(j)$ happened after item cluster $L(i)$.

\subsection{Compute Surprise Score}

Based on two relevance scores $s_1(i,j)$ and $s_2(i,j)$, we calculate the final related score by combining them linearly:
\begin{equation}\label{eq:ss}
s(i,j)=\omega*s_1(i,j) + (1-\omega)*s_2(i,j)
\end{equation}
where $\omega$ is the combination weight that can be set manually or estimated from the data. The default value of $\omega$ is 0.8 in our experiment unless otherwise noted.

\section{Experiments}

We will evaluate the proposed new approaches empirically. We first introduce the experimental setup, then report the offline evaluation with different metrics. The system is also deployed on Taobao.com to verify the results with online customers. Finally, efficiency analysis based on the running time is provided.

\subsection{Experimental Setup}

We collected a user behavior dataset with more than 400 million users and 500 million products from Dec 16th to Dec 30th 2015 in Taobao. The user click data is used to capture the substitute relationships, and the user purchasing data is used to capture the complementary relationships.

We utilize a classical item-based CF with cosine similarity as our main baseline method, as it fits well in the e-commerce scenario \cite{Linden:2003:ARI}.
Other similarity measures are also evaluated, and the cosine similarity performs best among existing choices.

For fair comparison, we also introduce a user weighting factor into the original CF to penalize the active users, which is similar to the way used in \textit{Swing} algorithm. This improvement further improves the performance of CF, and is used as a stronger baseline in this paper. Thus the final CF \textit{baseline method} is defined as follows.

\begin{displaymath}
w_{i,j}=\frac{\sum_{u \in U_i \bigcap U_j}w_u^2}{\sqrt{\sum_{u \in U_i}w_u^2} \sqrt{\sum_{v \in U_j}w_v^2}}
\end{displaymath}
\begin{displaymath}
w_{u} = \frac{1}{\sqrt{|I_{u}|}}, w_{v} = \frac{1}{\sqrt{|I_{v}|}}
\end{displaymath}
where $|I_{u}|$ is the product number of user clicks/purchases.

\subsection{Offline Evaluation}

\begin{figure}[h]\label{fig:offline}
\centering
\epsfig{file=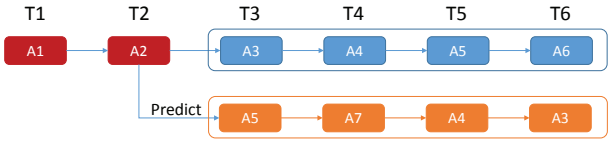, width=3.2in}
\caption{Offline evaluation example.}
\end{figure}

We propose an offline strategy to evaluate the performance of different techniques.
The basic idea is to see how the relations mined from historical data hit (i.e.~match) the real user behavior sequence in the future.
For example, as shown in figure~5, the top row is an actual user click sequence. We randomly choose an intermediate product in the sequence, such as $A2$ at time $T2$, as the seed product for substitute/similar products recommendation. The bottom row contains similar items recommended given product $A2$. In this case, the recommendations $hit$ $A3$, $A4$ and $A5$, the $hit$ number is 3. The larger the $hit$ number, the better the recommendation quality.

Given a set of product recommendation $predict_i$ of a given type (click or purchase) for a user $i$, and a set of known ground-$truth$, we can use traditional evaluation metrics, including Precision, Recall and MAP \cite{Baeza-Yates:1999}, for offline evaluation. The definitions of the metrics are:

\begin{displaymath}
 precision=\frac{1}{n}\sum_i^n {\frac{|hit_i|}{|predict_i|}}
\end{displaymath}

\begin{displaymath}
 recall=\frac{1}{n}\sum_i^n {\frac{|hit_i|}{|truth_i|}}
\end{displaymath}

\begin{displaymath}
 MAP=\frac{1}{n}\sum_i^n{\sum_{k=1}^{m} {precision_i@k}}
\end{displaymath}

where $hit_i=predict_i \cap truth_i$, and $predict_i$ denotes the algorithm prediction sequence for $i\mbox{-}th$ user, $truth_i$ denotes the real behavior (click/purchase) sequence. $precision_i@k$ denotes the precision of $predict_i$ with a cutoff at position $k$.

We utilize the data from 16th Dec to 30th Dec to mine the product relations with the proposed approaches and baseline methods, then we use the actual user behavior sequences generated from 31th Dec as the ground-truth\footnote{The dataset will be released later in our official website: https://tianchi.shuju.aliyun.com/datalab/index.htm}.
Evaluation results of $Swing$ are shown in Table~\ref{tab:sw}. The results show that the $Swing$ algorithm significantly outperforms the classical item-based CF approach. Specially, the relative improvement of Swing over CF is up to 67.6\% and 46.1\% in terms of Precision and Recall respectively. It indicates that our method could find more relevant substitute products with higher precision. Furthermore, our approach also outperforms CF in terms of MAP with orders of magnitude, which indicates relevant products are ranked higher in the recommendation lists by $Swing$.

The evaluation results for $Surprise$ are shown in Table~\ref{tab:sp}, and we have similar findings.

\begin{table*}[t]
\small \centering \caption{Offline evaluation of Swing.}
\label{tab:sw}
\begin{tabular*}{0.9\textwidth}{c||p{0.02\textwidth}p{0.25\textwidth}p{0.25\textwidth}p{0.25\textwidth}} \hline
~~~~~ && ~~~~Precision~~~~ & ~~~~Recall~~~~ & ~~~~~MAP \\\hline
CF & &~~~~~0.01471 & ~~~~0.1093  & ~~~~~0.01177 \\ \hline
Swing & &\textbf{0.02466}~~(+67.6\%)&	\textbf{0.1597}~~(+46.1\%)&	\textbf{0.06109}~~(+419\%) \\
\hline
\end{tabular*}
\end{table*}

\begin{table*}[t]
\small \centering \caption{Offline evaluation of Surprise.}
\label{tab:sp}
\begin{tabular*}{0.9\textwidth}{c||p{0.02\textwidth}p{0.25\textwidth}p{0.25\textwidth}p{0.25\textwidth}} \hline
~~~~~ && ~~~~Precision~~~~ & ~~~~Recall~~~~ & ~~~~~MAP \\\hline
CF & &~~~~~0.01188 & ~~~~0.1231  & ~~~~~0.06242 \\ \hline
Surprise & &\textbf{0.02519}~~(+111.9\%)&	\textbf{0.23875}~~(+93.9\%)&	\textbf{0.109558}~~(+75.5\%) \\
\hline
\end{tabular*}
\end{table*}

\subsection{Online Evaluation}

Online experiments with real world E-commerce users are carried out to study the effects of the proposed approaches. We conduct A/B tests based on recommendation scenarios in the mobile APP of Taobao.

The online metrics we utilize are Click Through Rate (CTR), Click Conversion Rate (CVR) and Payment Per Thousand Impressions (PPM), which are commonly used in e-commercial systems. The definitions are given as follows.

\begin{displaymath}
CTR=\frac{\#item\_{click}}{\#show\_pv}
\end{displaymath}

\begin{displaymath}
CVR=\frac{\#item\_trade}{\#item\_click}
\end{displaymath}

\begin{displaymath}
PPM=\frac{\#Payment}{\#show\_pv}*1000
\end{displaymath}

\subsubsection{Swing Online}

\begin{figure*}[!htb]
 \centering
  \subfigure[]{
    \label{fig:swo:a} 
    \includegraphics[width=0.32 \textwidth]{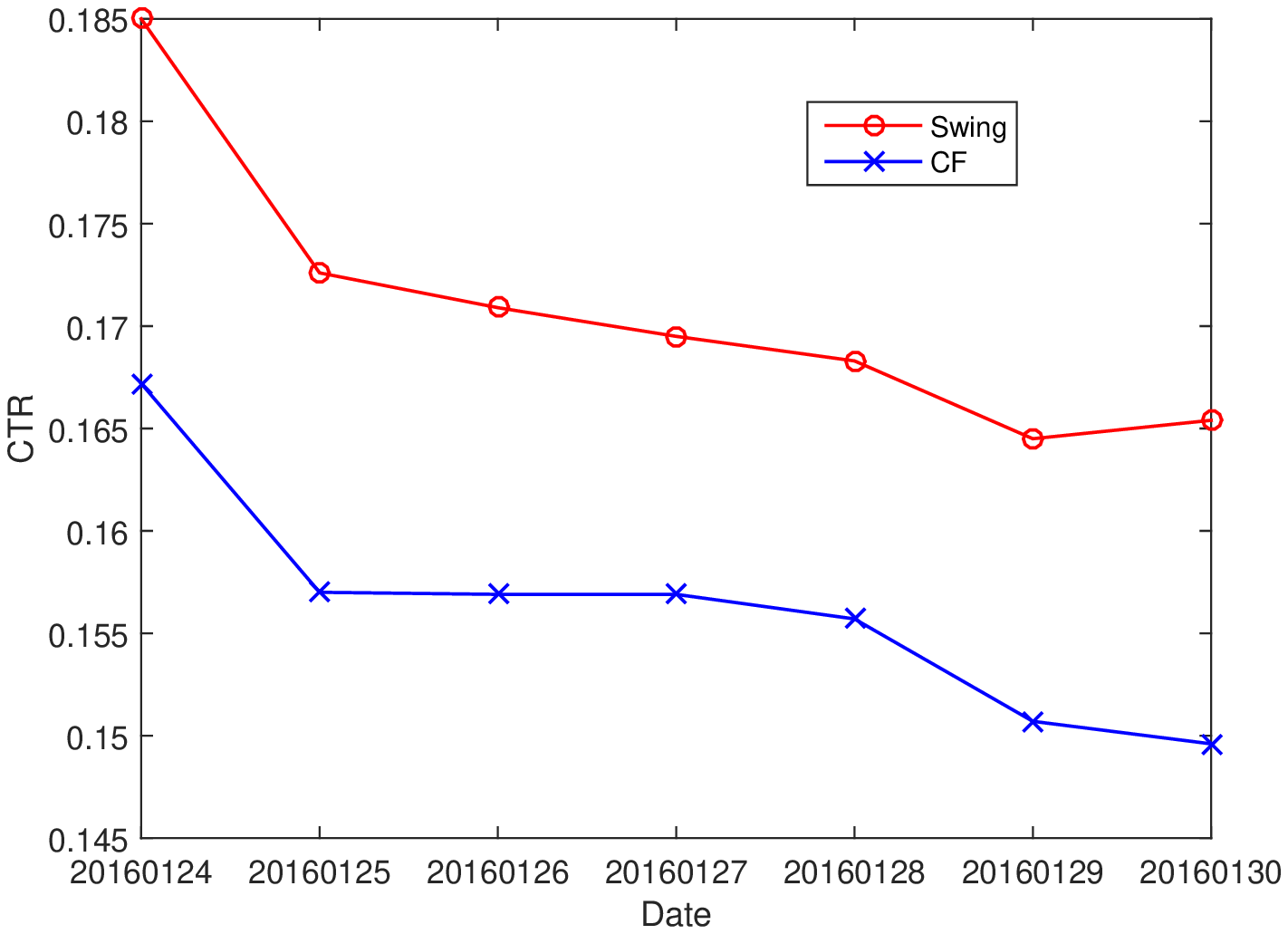}}
  \subfigure[]{
    \label{fig:swo:b} 
    \includegraphics[width=0.32 \textwidth]{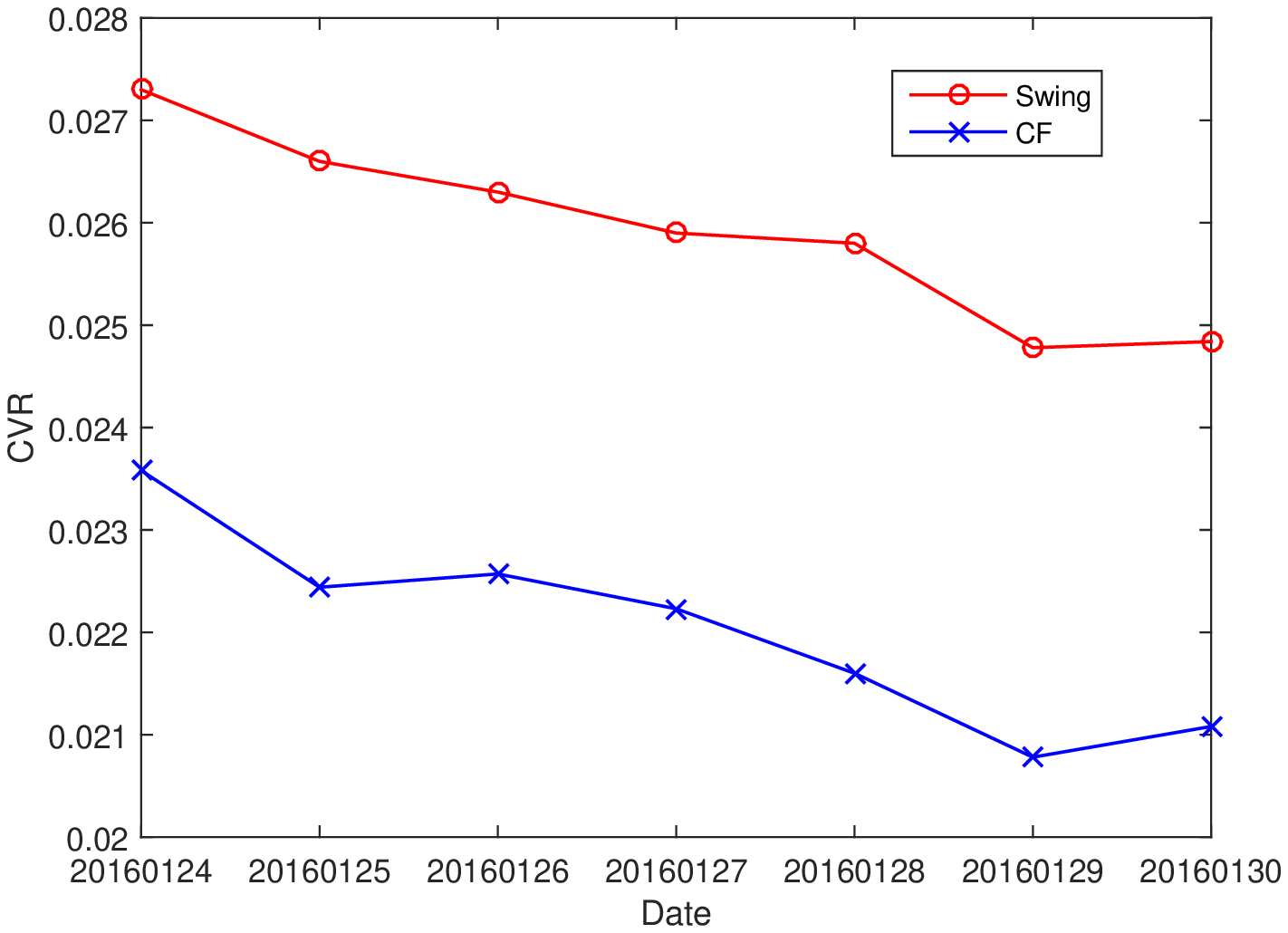}}
  \subfigure[]{
    \label{fig:swo:c} 
    \includegraphics[width=0.32 \textwidth]{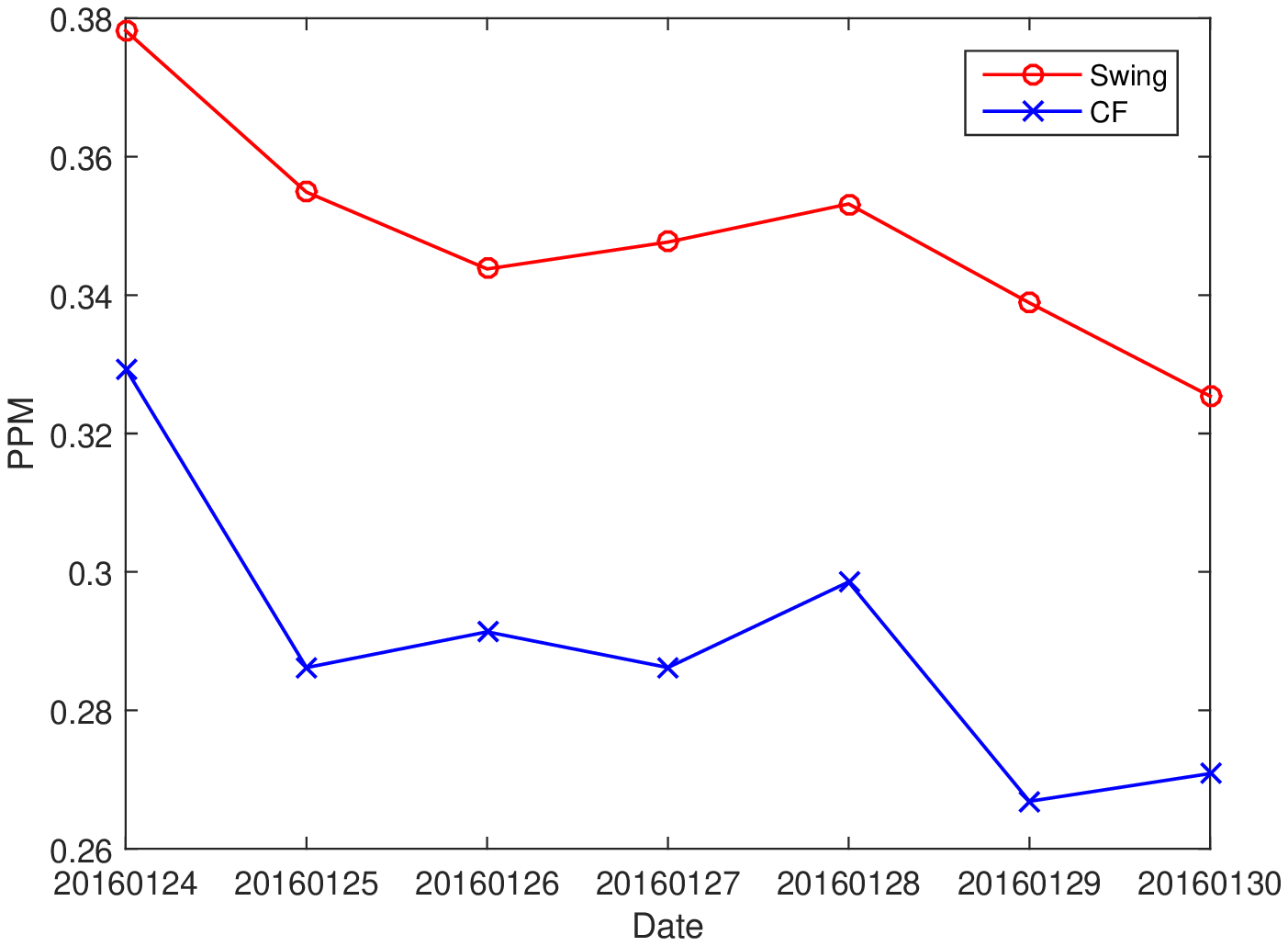}}
  \caption{Online Evaluation of Swing in pre-purchase scene via CTR, CVR and PPM metrics}
  \label{fig:swo}
\end{figure*}

We conducted A/B test with the $Swing$ algorithm and the baseline approach in pre-purchase scenarios, and the results are shown in Figure~\ref{fig:swo}. We find that the $Swing$ bucket significantly outperforms ($p-value<0.05$) the CF bucket in all three metrics. Specifically, the relative improvement of Swing over CF is up to 9.3\% and 17.6\% on average in terms of CTR and CVR, respectively. It means that users find more interested products with $Swing$, and the probability of converting the click behavior to real purchases is improved significantly. $Swing$ achieves an improvement of 20.3\% on average for PPM. PPM is the key criteria for recommendation in e-commerce, thus the improvement has a huge commercial value.

\subsubsection{Surprise Online}

\begin{figure*}[!htb]
 \centering
  \subfigure[]{
    \label{fig:spo:a} 
    \includegraphics[width=0.32 \textwidth]{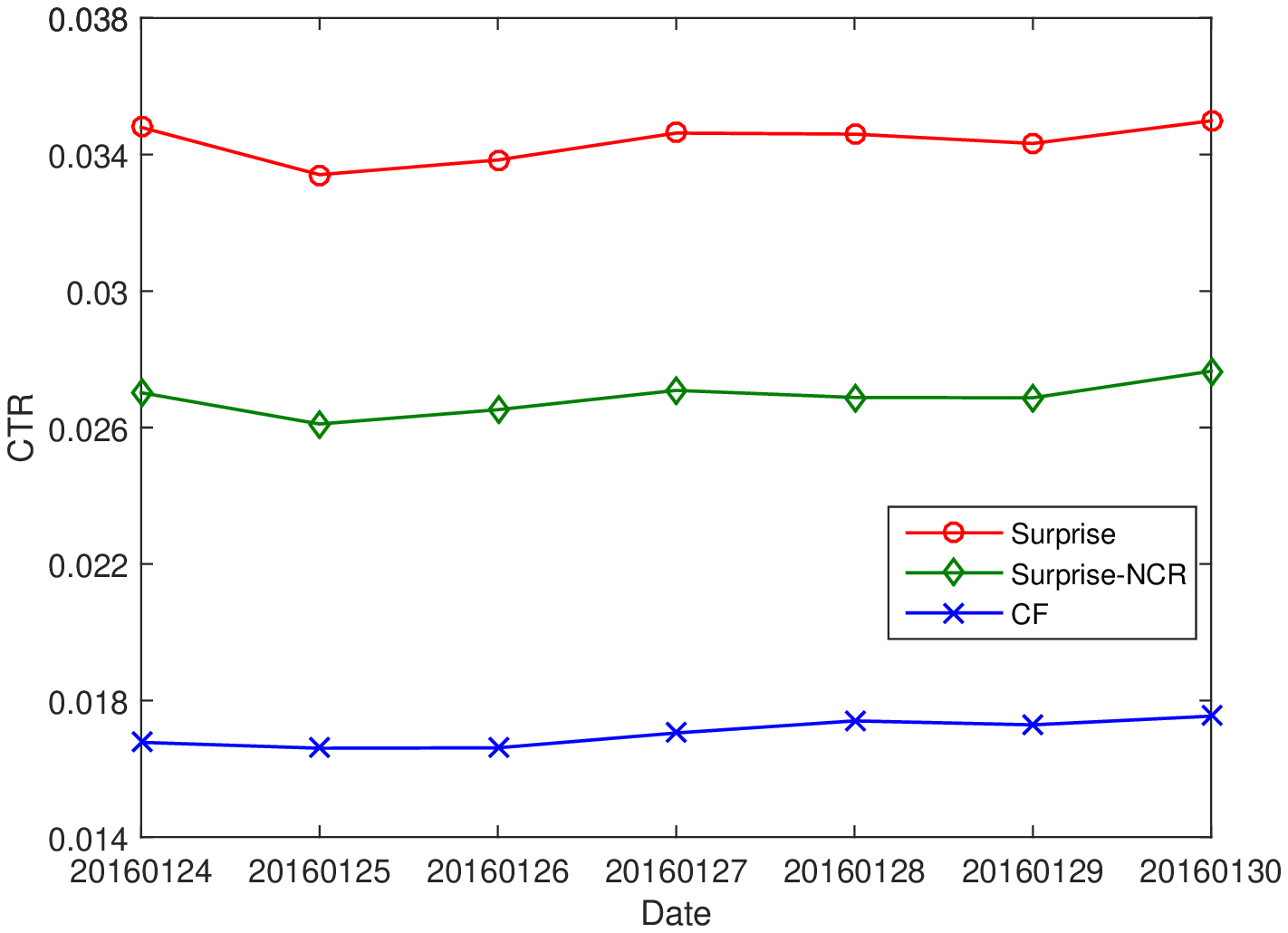}}
  \subfigure[]{
    \label{fig:spo:b} 
    \includegraphics[width=0.32 \textwidth]{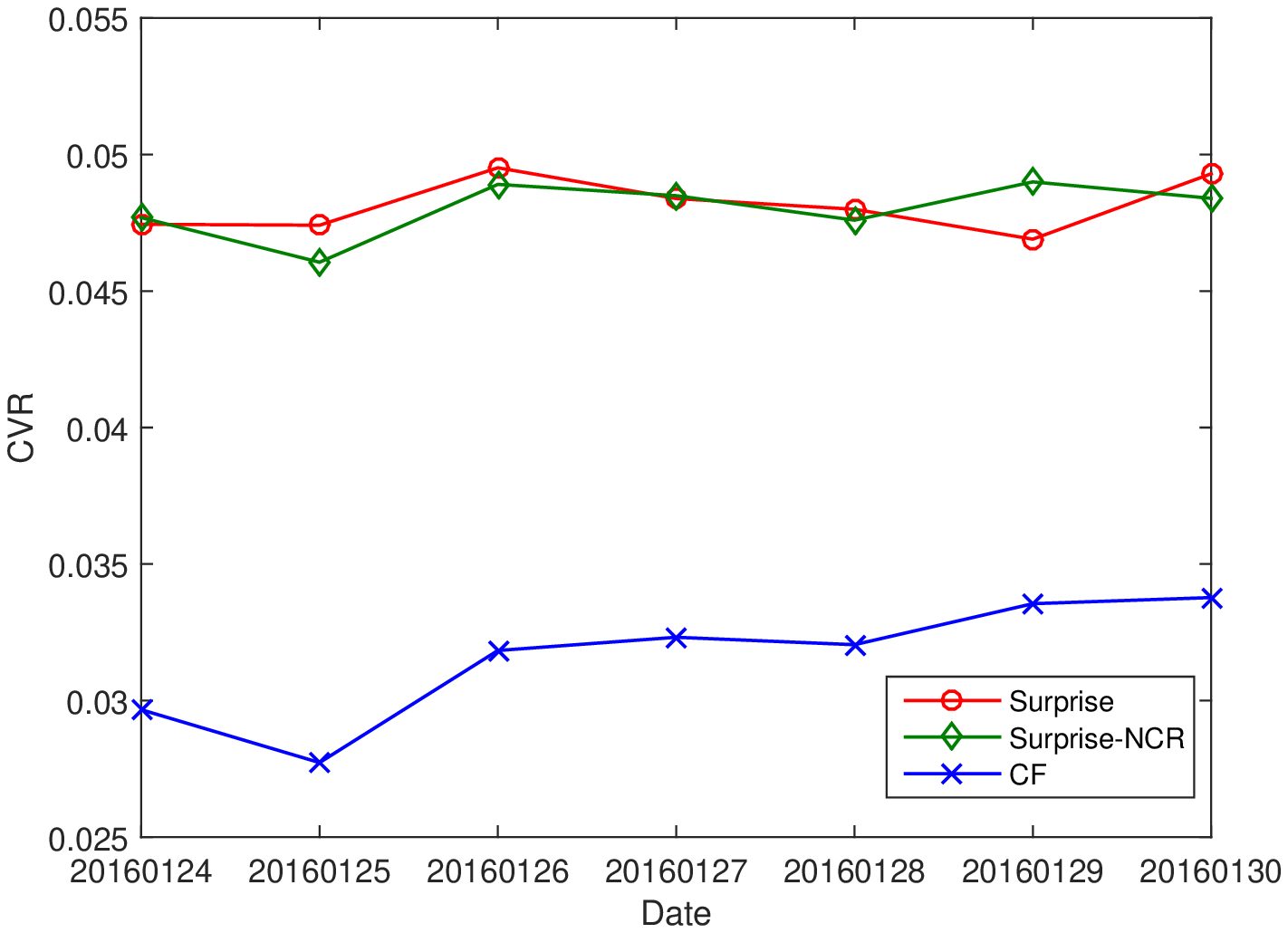}}
  \subfigure[]{
    \label{fig:spo:c} 
    \includegraphics[width=0.32 \textwidth]{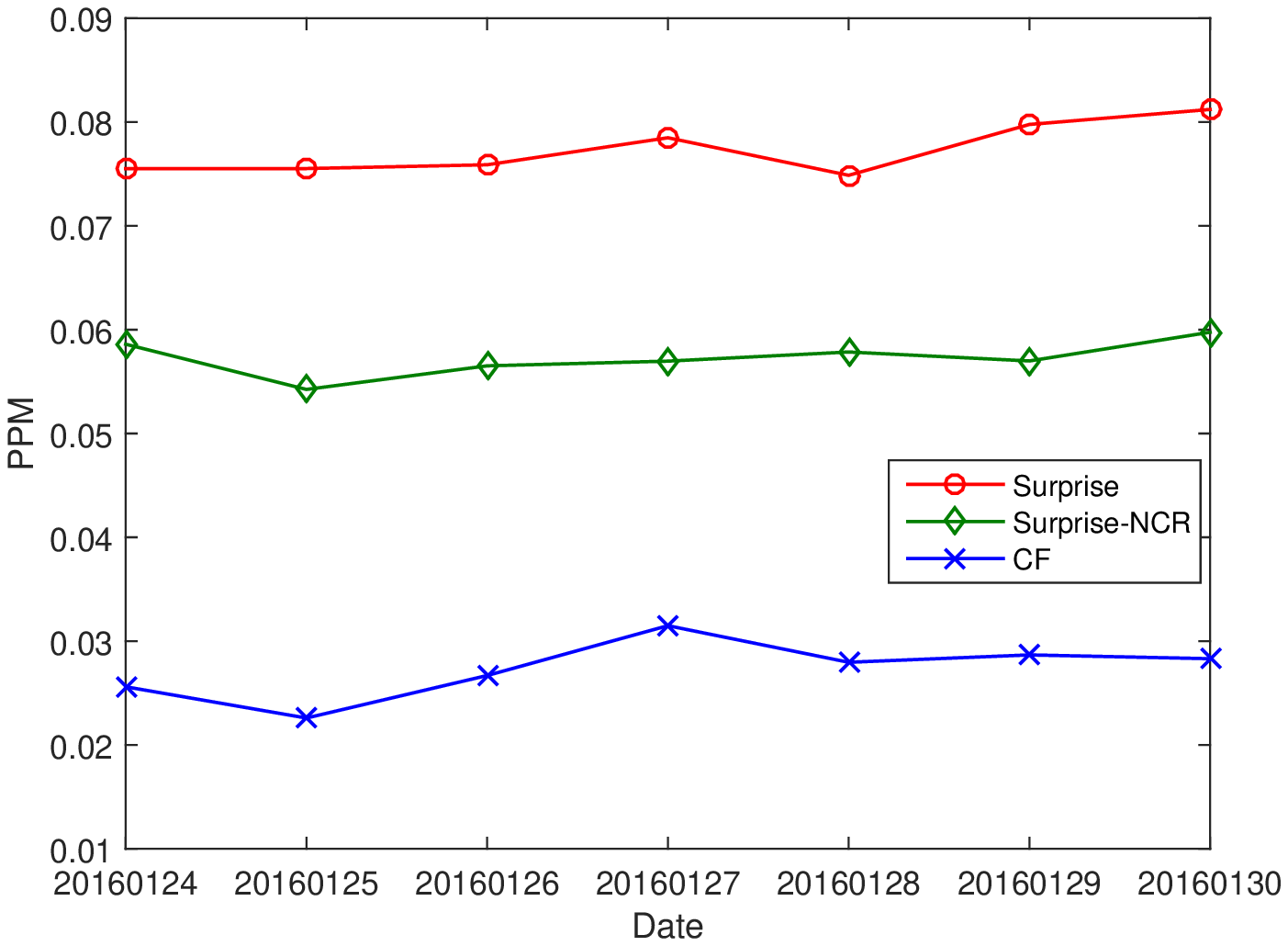}}
  \caption{Online Evaluation of Surprise in post-purchase scene via CTR, CVR and PPM metrics}
  \label{fig:spo}
\end{figure*}

The $Surprise$ algorithm is deployed in post-purchase scenarios, and the evaluation results are shown in Figure~\ref{fig:spo}.
For the $Surprise$ method, we also conducted another version without the relevance score at the cluster level, i.e. $s(i,j)=s_1(i,j)$, and it is represented as $Surprise\mbox{-}NCR$.

From the figures we can find that both $Surprise$ and $Surprise\mbox{-}NCR$ show significant improvements over the original purchasing based CF in all evaluation measures, which indicates the importance of top-related categories and the time-based weighting mechanism.
Meanwhile, compared with $Surprise\mbox{-}NCR$, $Surprise$ achieved 27.9\% CTR improvement on average, largely due to better relevance estimation and its ability to generate more product recommendations. On the CVR metric, $Surprise\mbox{-}NCR$ has very similar performance to Surprise. Our hypothesis is that once a user has clicked a certain interested product, the willingness to purchase is largely depends on the product properties itself, which is not captured by the cluster-level relevance score. With the measure of PPM, $Surprise$ outperforms $Surprise\mbox{-}NCR$ and the original purchasing CF by 35\% and 183\% on average, respectively. In summary, the $Surprise$ algorithm achieved the best commercial impact.

\subsubsection{Hybrid Application}

\begin{figure*}[!htb]
 \centering
  \subfigure[]{
    \label{fig:hy:a} 
    \includegraphics[width=0.32 \textwidth]{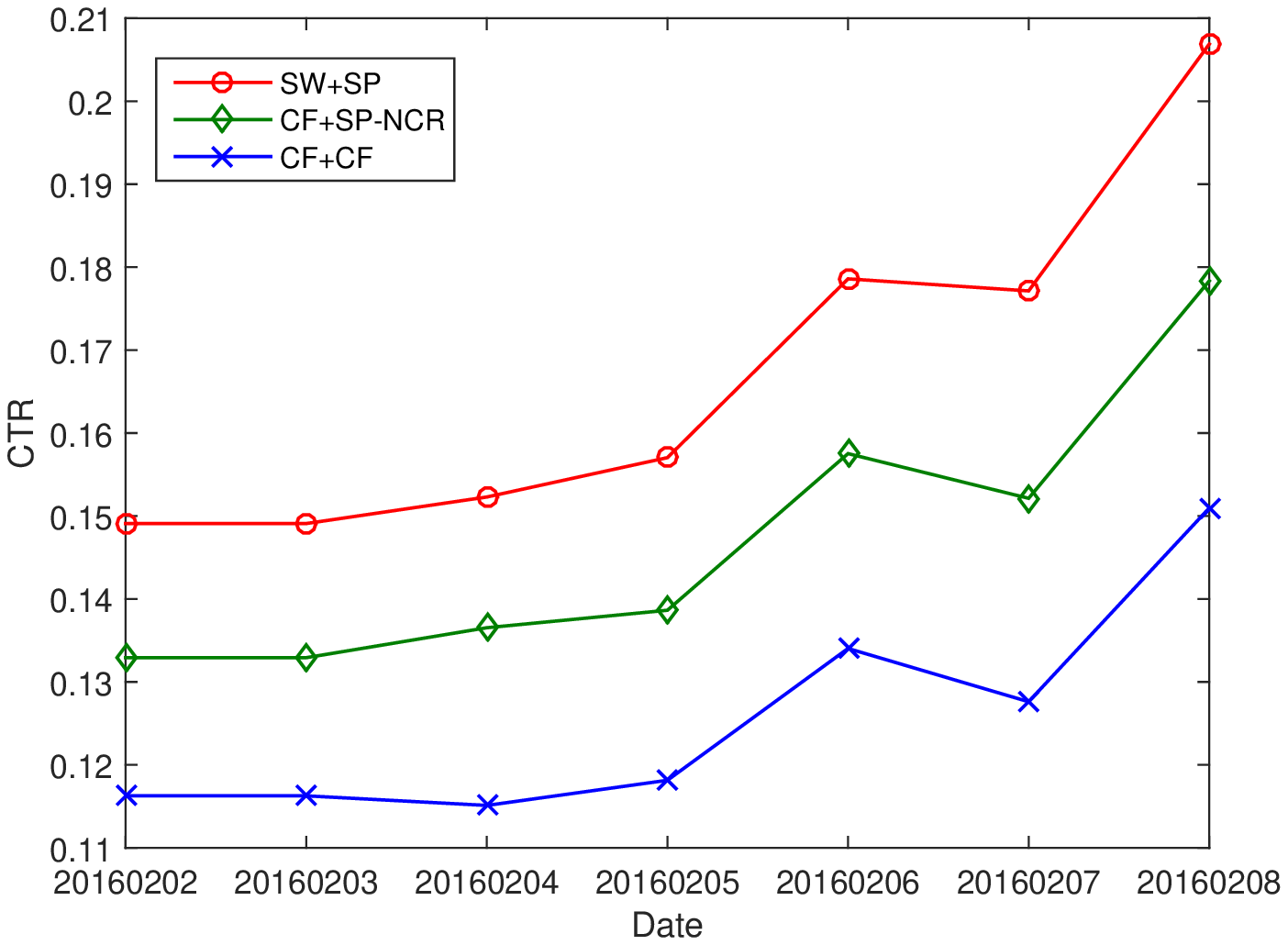}}
  \subfigure[]{
    \label{fig:hy:b} 
    \includegraphics[width=0.32 \textwidth]{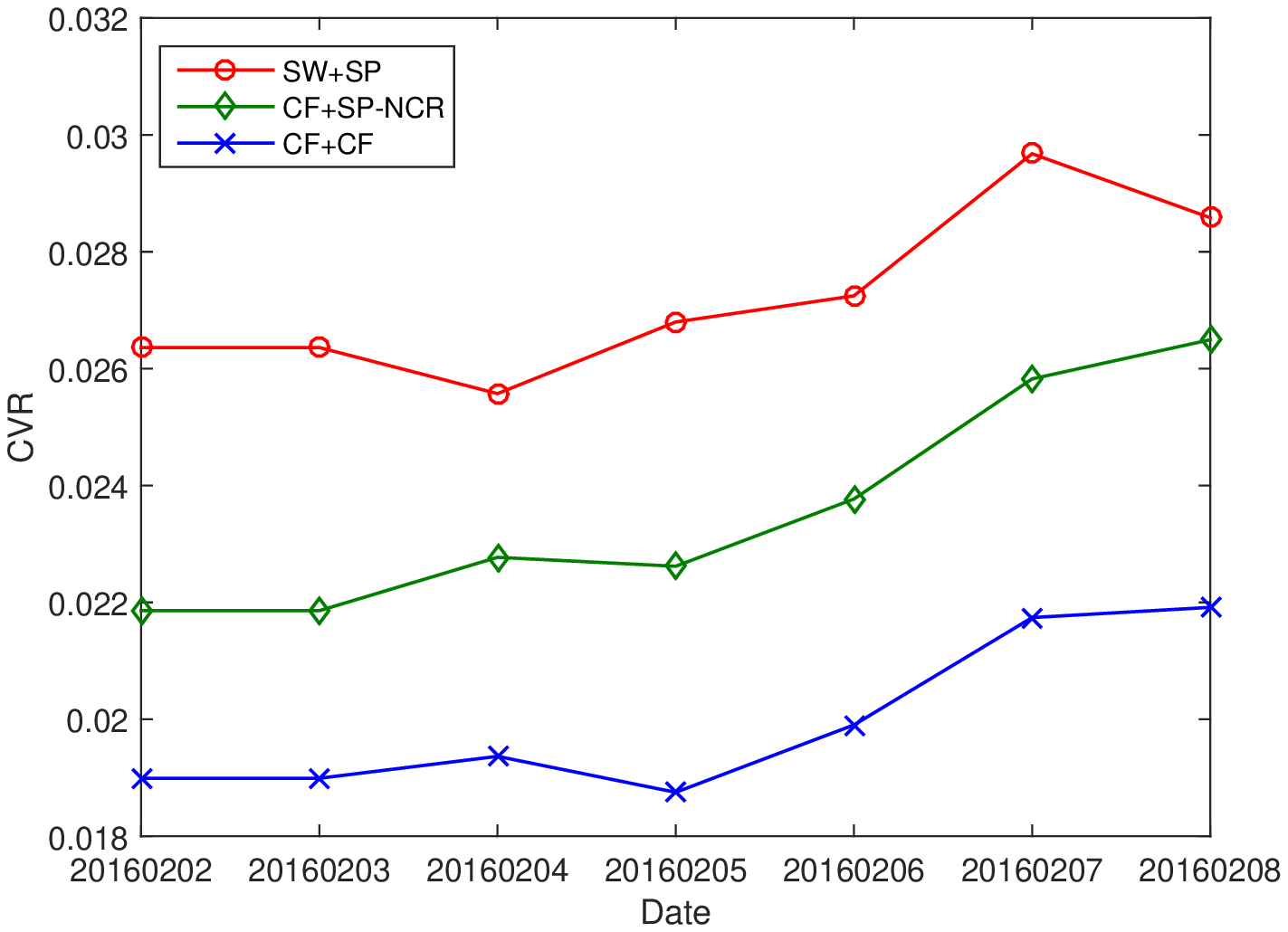}}
  \subfigure[]{
    \label{fig:hy:c} 
    \includegraphics[width=0.32 \textwidth]{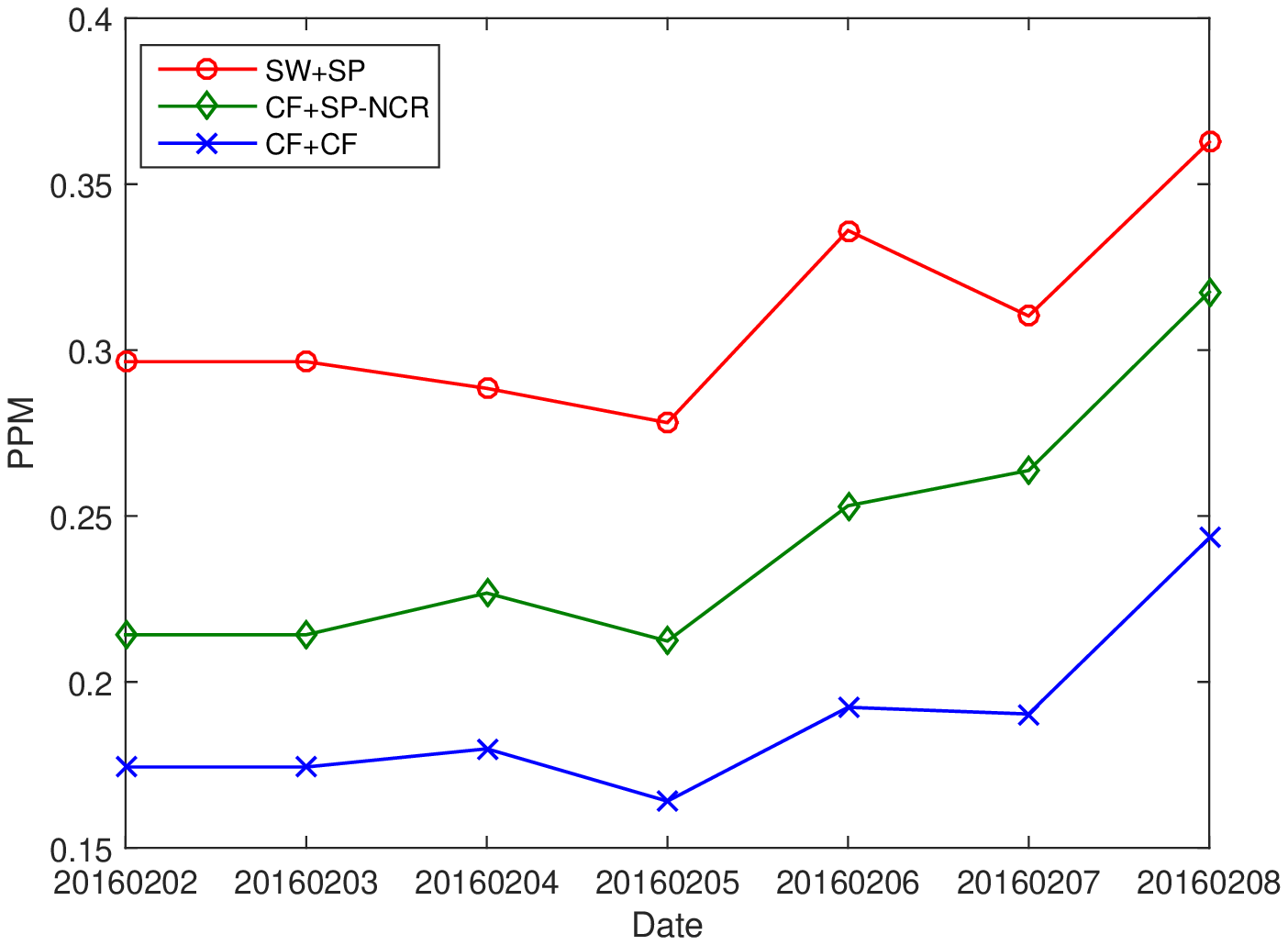}}
    \vspace{-0.2cm}
  \caption{Online Evaluation of Hybrid application in Taobao via CTR, CVR and PPM metrics}
  \vspace{-0.2cm}
  \label{fig:hy}
\end{figure*}

We also conduct an overall evaluation of the product graphs in an integrated scenario, which are constructed by Swing (SW) and Surprise(SP).
The recommendation list is generated by combing substitute items for user past clicks and complementary items for user current order. The detailed evaluation results are shown in Figure~\ref{fig:hy}. The combination of our proposed methods outperforms the original CF + CF by 33.2\%, 26.7\%, 62.9\% on average in CTR, CVR and PPM, respectively. The results further prove the effectiveness of our proposed approaches.

\subsection{Efficiency Evaluation}

We further study the efficiency of the proposed approaches. As discussed in section~\ref{ss:sw}, the time complexity of Swing is higher than CF due to the consideration of inner structures on user click graph. While for both methods, we developed practical parallelization implementation for large scale applications, it is still worth to compare the offline calculation time of all the methods. The results are shown as following (unit: hour):
\begin{displaymath}
\text{Click-CF}~(\sim 2.0h) \prec \text{Swing}~(\sim 3.5h)
\end{displaymath}
\begin{displaymath}
\text{Purchase-CF}~(\sim 1.3h) \prec \text{Surprise}~(\sim 2.5h)
\end{displaymath}

The computation of Surprise includes 3 parts: the inference of top-related categories, the item-level relevance calculation and the cluster-level relevance calculating. With top relevant categories, we can pre-filter unrelated items at the Mapper stage and reduce the calculation cost greatly.
Our approach takes longer, however, acceptable offline computing time, while achieves significant improvements in online experiments.

\section{Related Work}

Recommender systems can help consumers to select the right products to match their tastes and meet their special needs, and play a key role in modern e-commerce systems. Most of existing recommendation approaches can be divided into two categories: Content-Based approaches and Collaborative Filtering approaches.

The content-based approaches make recommendation based on item features. Several approaches have been proposed \cite{Mooney:2000:CBR, Pazzani:1997, Lops:2011}, such as vector space models, bayesian classifiers and clustering models. However, the application of CB methods in e-commerce is limited, because content features such as product title words, brands can hardly describe all the important hidden characters of products and the fine-grained preferences of users. Therefore, CB methods are usually utilized as part of a solution in a hybrid recommender system \cite{Burke:2002:HRS, Wang:2011:URP}.

The collaborative Filtering approaches make recommendations based on a user's past behaviors \cite{Linden:2003:ARI, Wang:2011:URP}. These approaches usually operate on the user-item matrix, where each row is a user vector and each column is a item vector. User-based methods try to find similar user neighbors to the current user \cite{Sarwar:2000:ARA, Breese:1998:EAP}. Item-based approaches find item neighbors that are similar to the item a user clicked or purchased based on the item vectors \cite{Sarwar:2001:ICF, Karypis:2001:EIT, Deshpande:2004:ITN}. Among these methods, various similarity measures have been proposed to find the top-n similar neighbors, such as cosine similarity, Jaccard similarity, Pearson correlation coefficient and conditional probability based similarity.
Compared with the user-based approaches, the item-based methods are more commonly used in large scale e-commerce systems due to their effectiveness and efficiency.
Recently, factorization models such as Singular Value Decomposition (SVD) \cite{Koren:2009:MFT} or Probabilistic Matrix Factorization (PMF) \cite{Salakhutdinov:2008:BPM} have gained much attention due to their good performance on some benchmark datasets, and those methods can also incorporate additional information such as implicit user feedback and temporal information.

Rather than providing a single recommendation algorithm or solution, this paper mainly focuses on finding the relationships between products, which serve as the basis for quickly returning candidate products for further computationally expensive recommendation algorithms, such as hybrid filtering or learning to rank. A recent work \cite{McAuley:2015:INS} also focuses on the same problem. It utilizes the text of product reviews and descriptions to infer the relationships between products via a supervised approach. In contrast to their work, we focus on directly utilizing the user behavior data such as user co-click and co-purchase, which are stronger and more reliable signals for the capturing of product relationships.

Although we are constructing item-item similarity matrix, traditional item-based CF methods based on standard local similarity measures \cite{Sarwar:2001:ICF, Linden:2003:ARI, zhou2009predicting} are `pointwise' measures, which ignores the inner structure of user behavior data. On the other hand, existing work on social network and graph analysis have shown inner structure could be useful for predictions in other domains. These motivate us to introduce inner structures to recommendation system.

\section{Conclusions}

This paper focus on the product graph with substitute and complementary relationships.
Construction of such kind of large scale product graphes in e-commerce has several major challenges: robustness, data noise, data sparsity, relationship directions and scalability. To tackle those challenges, we propose the Swing algorithm, which can utilize the inner stable structures of user behavior data, to capture the substitute relationships of products. It performs significantly better as it eliminates noisy information and makes the predicted relationship more robust.
Then we propose an advanced Surprise algorithm for the modeling of complementary relationship. Surprise utilizes product category information and solves the sparsity problem in user co-purchasing graph via clustering technique of label propagation. Furthermore, it also considers the time sensitivity and time sequential of propagation to guarantee the complementary relationship is reasonable. Base on the two approaches, we build two product graphes to support fast recommendations in Taobao. Finally, we verify the effectiveness and efficiency of our approaches with comprehensive offline and online evaluations.

The product-product graph construction methods proposed in this paper is very general and can be applied to other applications, such as advertising and personalized search in e-commerce. This can be explored in the future. Besides, we will consider combing the content information with the proposed work more effectively and efficiently. This paper is a first step towards introducing quasi-local user-user edge based inner structure into recommendation systems. The approach is simple, efficient and extremely effective in practice. We plan to consider different inner structures and introduce them in other ways such as Markov random fields or Bayesian network in the future, and explore efficient solutions to adapt those more theoretical solutions to our large scale system.

%
\bibliographystyle{abbrv}
\bibliography{sigproc}  
%
%

\end{document}